\documentclass[aps,pra,10pt,groupedaddress,showpacs]{revtex4-1} 

\usepackage{amsmath}
\usepackage{graphicx}
\usepackage{amsfonts}
\usepackage{color}
\usepackage{amssymb}
\usepackage{mathrsfs}
\usepackage{hyperref}
\usepackage{color}

\newcommand{\ket}[1]{\left|#1\right>} 
\newcommand{\bra}[1]{\left<#1\right|} 
\newcommand{\avg}[1]{\left< #1\right>}
\newcommand{\tr}{\mathrm{tr}}
\newcommand{\rme}{\mathrm{e}}
\newcommand{\rmi}{\mathrm{i}}
\newcommand{\rmd}{\mathrm{d}}
\newcommand{\etal}{\textit{et al.}}

\begin{document}

\title{\textcolor{black}{Loss-tolerant hybrid measurement test of CHSH inequality with weakly amplified N00N states}}
\author{Falk T\"{o}ppel}
\email[]{falk.toeppel@mpl.mpg.de} 
\affiliation{Max Planck Institute for the Science of Light, G\"{u}nther-Scharowsky-Stra{\ss}e 1/Bldg. 24, 91058 Erlangen, Germany}
\affiliation{Institute for Optics, Information and Photonics, Universit\"{a}t Erlangen-N\"{u}rnberg, Staudtstra{\ss}e 7/B2, 91058 Erlangen, Germany}
\affiliation{Erlangen Graduate School in Advanced Optical Technologies (SAOT), Paul-Gordan-Stra{\ss}e 6, 91052 Erlangen, Germany}
\author{Magdalena Stobi\'nska}
\affiliation{Institute of Theoretical Physics and Astrophysics, University of Gda\'nsk, ul. Wita Stwosza 57, 80-952 Gda\'nsk, Poland}
\affiliation{Institute of Physics, Polish Academy of Sciences, Al. Lotnik\'ow 32/46, 02-668 Warsaw, Poland}
\date{\today}

\begin{abstract}
Although our understanding of Bell's theorem and experimental techniques to test it have improved over the last 40 years, thus far all Bell tests have suffered at least from the detection or the locality loophole. Most photonic Bell tests rely on inefficient discrete-outcome measurements, often provided by photon counting detection. One possible way to close the detection loophole in photonic Bell tests is to involve efficient continuous-variable measurements instead, such as homodyne detection. Here, we propose a test of the Clauser-Horne-Shimony-Holt (CHSH) inequality that applies photon counting and homodyne detection on weakly amplified two-photon N00N states. The scheme suggested is remarkably robust against experimental imperfections and suits the limits of current technology. As amplified quantum states are considered, our work also contributes to the exploration of entangled macroscopic quantum systems. Further, it may constitute an alternative platform for a loophole-free Bell test, which is also important for quantum-technological applications.
\end{abstract}


\maketitle

\section{Introduction}

Bell's theorem states that no local realistic theory can reproduce all the predictions of quantum mechanics~\cite{Bell}. It can be expressed in the form of various inequalities which are violated by some entangled states.
These Bell's inequalities resulted from a philosophical debate addressing fundamental questions on the description of physical phenomena~\cite{Bell,EPR}. 
Lately however, practical perspectives have been added to the discussion since Bell's inequality can also be viewed as a tool for the development of quantum technologies~\cite{Brukner}. The violation of a Bell's inequality certifies quantum correlations allowing for: device-independent quantum key distribution (QKD)~\cite{Acin2007}, verification of the security of QKD protocols~\cite{Scarani}, randomness generation~\cite{Pironio2010} and reduction of communication complexity~\cite{Zukowski}, to name just a few examples. Therefore, it is desirable to demonstrate violation of Bell's inequality for various quantum states in diverse configurations.

Violation of Bell's inequalities has been observed in many experiments using light~\cite{Aspect}, atoms~\cite{Hofmann}, ions~\cite{Rowe} and superconducting electric circuits~\cite{Ansmann}. However, all experiments performed thus far have suffered at least from either the detection or the locality loophole.
Entangled pairs of photons allow for a space-like separation between the measurements fairly easily, but the efficient quantum photon detectors (detection loophole) are missing; quantum states of matter systems can be tested efficiently, but a space-like separation (locality loophole) is hard to achieve in experiment.
The challenge is to close both loopholes simultaneously in one experiment.

Recently, the groups of A. Zeilinger~\cite{Giustina} and of P. G. Kwiat~\cite{Christensen} claimed to have closed the detection loophole for the first time in an experimental test of Bell's inequality with light. These experiments make photons the first system with both loopholes closed, although not in the very same experiment.
Most of the Bell tests, including those discussed in references~\cite{Giustina} and~\cite{Christensen}, rely on pairs of photons in a singlet state and photon counting, i.e. discrete variable measurements, which used to be not very efficient. Advances in detector technology, however, namely the development of superconducting transition edge sensors~\cite{detector1,detector2,detector3}, enabled the detection loophole to be closed.

A different way to perform an optical Bell test free of the detection loophole is by involving efficient continuous-variable measurements, e.g., homodyne detection~\cite{Gilchrist}. The difficulty here lies in finding non-Gaussian quantum states that violate Bell's inequality and are easy to prepare. As promising alternative, hybrid schemes that use continuous and discrete variable measurements have been proposed to implement Bell tests with entangled photons~\cite{2photonbell,Brask,Quintino} or atom-photon entanglement~\cite{PhysRevA.84.052122,Teo}. \textcolor{black}{For certain, possibly infeasible states, these schemes, in principle, allow violations of Bell's inequality at arbitrary low detection efficiencies \cite{PhysRevA.86.030101}.}

Another approach to tackle the detection loophole is to work with macroscopic quantum states of light~\cite{Vitelli,Stobinska2011,Sekatski}. These states are produced experimentally by amplifying vacuum or a few-photon quantum states with an optical parametric amplifier (OPA). In this process, the quantum properties of the initial state are preserved such that, for example, macroscopic entanglement can be verified~\cite{demartini,Iskhakov,BSV}. As macroscopic quantum states contain many photons, they are registered with high probability despite low detection efficiencies. Moreover, subjecting macroscopic quantum states of light to quantum engineering can facilitate using detectors with finite resolution, worse than single-photon resolution~\cite{MDF,CPC,Feasibility}. 

A link between Bell tests and macroscopic quantum states are entangled two-photon N00N states $\ket{\psi_2}$, where $\ket{\psi_N}=(1/\sqrt{2})(\ket{N}_A\!\ket{0}_B+\rme^{\rmi\varphi}\ket{0}_A\!\ket{N}_B)$ with $A$ and $B$ denoting different modes, e.g. orthogonal polarization. Two-photon N00N states have been proposed for a violation of Bell's inequality in a hybrid scheme~\cite{2photonbell} and have been amplified to macroscopic scales~\cite{production}. In this article, we join both results and establish a potentially loophole-free Bell test, using discrete and continuous-variable measurements, for amplified two-photon N00N states. We find that weak amplification yields a bigger violation of Bell's inequality than that achieved in~\cite{2photonbell} for the unamplified state. Our proposal is remarkably robust against experimental imperfections and suits the limits of current technology. 
In addition, we underline with our work the opportunities and explore the limits of entangled macroscopic quantum states for quantum information processing.

\section{Amplified two-photon N00N state}
\label{sec:AN00N}
The amplification of an arbitrary two-mode quantum state to macroscopic scales is usually performed with an OPA~\cite{demartini,spagnolo}. Its action on an input quantum state is described by the unitary evolution operator $\hat{S}_A(\zeta)\hat{S}_B(\zeta)$, where $A$ and $B$ label orthogonal modes, e.g. polarization modes. Here, $\hat{S}_A(\zeta_A)=\exp\left[\frac{1}{2}(\zeta_A^*\hat{a}^2-\zeta_A\hat{a}^{\dagger 2})\right]$ and $\hat{S}_B(\zeta_B)$, defined accordingly, denote the single-mode squeezing operator for mode $A$ and $B$. Thus, the amplification of a two-photon N00N state $\ket{\psi_2}$ results in the quantum state~\cite{production}
\begin{subequations}
\begin{align}
\label{eq:an00n_state}
\ket{\Psi_2}&=\hat{S}_A(\zeta_A)\hat{S}_B(\zeta_B)\ket{\psi_2}\\
&=\frac{1}{\sqrt{2}}\hat{S}_A(\zeta_A)\hat{S}_B(\zeta_B)\left(\ket{2}_A\!\ket{0}_B+\rme^{\rmi\varphi}\ket{0}_A\!\ket{2}_B\right).\nonumber
\end{align} 
In the following, we will only consider the case $\zeta_A=\zeta_B=\zeta\geq0$ and $\varphi=0$. 
Let us express $\ket{\Psi_2}$ in the photon number state representation. To this end, we rewrite Eq.~(\ref{eq:an00n_state}) as 
\begin{align}
\label{eq:an00n_state_b}
\ket{\Psi_2}=\frac{1}{\sqrt{2}}(\ket{\Phi_2}_A\ket{\Phi_0}_B+\ket{\Phi_0}_A\ket{\Phi_2}_B), 
\end{align}
\end{subequations}
with $\ket{\Phi_0}=\hat{S}(\zeta)\ket{0}$ and $\ket{\Phi_2}=\hat{S}(\zeta)\ket{2}$ being the squeezed vacuum and squeezed two-photon state~\cite{Kral}:
\begin{align*}
\ket{\Phi_0}&=\frac{1}{\sqrt{\mu}}\sum_{n=0}^\infty\left[-\frac{\nu}{2\mu}\right]^n\!\frac{\sqrt{(2n)!}}{n!}\ket{2n},\\
\ket{\Phi_2}&=\frac{1}{\sqrt{2\mu^3}}\sum_{n=0}^\infty\left[-\frac{\nu}{2\mu}\right]^n\!\frac{\sqrt{(2n)!}}{n!}\!\left[\nu-\frac{2n}{\nu}\right]\!\ket{2n},
\end{align*}
where $\nu=\sinh|\zeta|$ and $\mu=\cosh|\zeta|$.

The mean total photon number of the amplified N00N states, $\bar{n}_\mathrm{tot}=\bra{\Psi_2}(\hat{n}_A+\hat{n}_B)\ket{\Psi_2}=2+6\sinh^2|\zeta|$, is determined by the parametric gain $\zeta$ of the OPA and can reach macroscopic scales. Nevertheless, the states consist of components with only even photon numbers, a typical feature of non-classical single-mode squeezed states.

\section{Violation of Bell's inequality with amplified two-photon N00N states}
\label{sec:bell_test}
Cavalcanti \etal~\cite{2photonbell} suggest a Bell test that violates the Clauser-Horne-Shimony-Holt (CHSH) inequality \cite{CHSH} using the two-photon N00N state $\ket{\psi_2}$. Since $\ket{\Psi_2}$ is the macroscopic version of that state, we study the advantages of using amplified two-photon N00N state in the same setup. The proposed experiment is performed by two spatially separated observers Alice ($A$) and Bob ($B$). Each observer can decide whether he measures the photon number ($N$) of the incident beam or performs homodyne detection ($X$). Depending on the choice and the measurement outcome, Alice (Bob) assigns $+1$ or $-1$ to the observables $b_A(N)$ and $b_A(X)$ ($b_B(N)$ and $b_B(X)$). More precisely, after having performed a photon number measurement, the observer $j\in\{A,B\}$ assigns $+1$ to  $b_j(N)$ if the photon number obtained $n_j$ is below a threshold $n_0$ and $b_j(N)=-1$ otherwise. Similarly, when the observer $j$ performs homodyne detection, $+1$ is assigned to $b_j(X)$ for a quadrature value $x_j$ with modulus greater than $x_0$ and $-1$ otherwise. In summary, this protocol reads for $m\in\{N,X\}$ as:
\begin{align}
\label{eq:definition_bell}
b_j(m)=\left\{\begin{array}{ll}
  -1, & \mathrm{if~}[m=N\mathrm{~and~}n_j > n_0]\mathrm{~or~}[m=X\mathrm{~and~}|x_j|\leq x_0],\\
  +1, & \mathrm{if~}[m=N\mathrm{~and~}n_j \leq n_0]\mathrm{~or~}[m=X\mathrm{~and~}|x_j|> x_0].
\end{array}\right.
\end{align}
Since the measurement described above yields dichotomic outcomes, the CHSH inequality can be applied~\cite{2photonbell}:
\begin{align}
\label{eq:mean_val_Bell}
\overline{\mathcal{B}}=\overline{E}_{XX}+\overline{E}_{XN}+\overline{E}_{NX}-\overline{E}_{NN}\leq2.
\end{align}
Therein 
\begin{align*}
\overline{E}_{mm'}&=\overline{b_A(m)b_B(m')}
=P[b_A(m)=b_B(m')]-P[b_A(m)\neq b_B(m')],
\end{align*}
denotes the correlation between the measurement outcomes when Alice and Bob perform the measurements $m\in\{N,X\}$ and $m'\in\{N,X\}$, respectively. The CHSH inequality (\ref{eq:mean_val_Bell}) is associated with a Bell observable $\hat{\mathcal{B}}$ such that $\overline{\mathcal{B}}=\mathrm{tr}\{\hat{\rho}\hat{\mathcal{B}}\}$ with $\hat{\rho}$ as the quantum state considered for the Bell test. In the following, the explicit form of this Bell observable is derived.

The detection of $n$ photons in an ideal quantum photon detector is described by the projector $\ket{n}\!\bra{n}$, where $\ket{n}$ denotes the eigenstate of the photon number operator $\hat{n}$. Likewise, the measurement of the quadrature value $x$ in perfect homodyne detection is represented by a projection on the eigenstate $\ket{x}$ of the quadrature operator $\hat{x}=(\hat{a}+\hat{a}^\dagger)/2$. In photon number state representation, it is
\begin{align*}
\ket{x}=\left(\frac{2}{\pi}\right)^{1/4}\sum_{n=0}^\infty\frac{\rme^{-x^2}}{\sqrt{2^n n!}}\,\mathrm{H}_n(\sqrt{2}x)\ket{n},
\end{align*}
with $\mathrm{H}_n(x)$ denoting the Hermite polynomial of order $n$. 
Due to the identities $\sum_{n=0}^\infty\ket{n}\!\bra{n}=1$ and $\int_{-\infty}^\infty \rmd x\ket{x}\!\bra{x}=1$, we can express the condition given in Eq.~(\ref{eq:definition_bell}) in terms of projectors
\begin{align*}
\hat{\Pi}_{N,j}^+&=\sum_{n=0}^{n_0}\ket{n}\!\bra{n}_j,\\
\hat{\Pi}_{N,j}^-&=1-\sum_{n=0}^{n_0}\ket{n}\!\bra{n}_j,
\end{align*}
corresponding to the photon number measurement of the observer $j\in\{A,B\}$ and the projectors
\begin{align*}
\hat{\Pi}_{X,j}^+&=1-\int_{-x_0}^{x_0} \rmd x\ket{x}\!\bra{x}_j,\\
\hat{\Pi}_{X,j}^-&=\int_{-x_0}^{x_0} \rmd x\ket{x}\!\bra{x}_j,
\end{align*}
%
referring to the homodyne measurement. 
Hence, we find the following operational form of the Bell observable:
\begin{align}
\label{eq:bell_observable}
\hat{\mathcal{B}}=\hat{E}_{XX}+\hat{E}_{XN}+\hat{E}_{NX}-\hat{E}_{NN},
\end{align}
where 
\begin{subequations}
\label{eq:bell_correlations}
\begin{align}
\hat{E}_{XX}&=\bigl[\hat{\Pi}_{X,A}^+-\hat{\Pi}_{X,A}^-\bigr]\!\otimes\bigl[\hat{\Pi}_{X,B}^+-\hat{\Pi}_{X,B}^-\bigr],\\ 
\hat{E}_{XN}&=\bigl[\hat{\Pi}_{X,A}^+-\hat{\Pi}_{X,A}^-\bigr]\!\otimes\bigl[\hat{\Pi}_{N,B}^+-\hat{\Pi}_{N,B}^-\bigr],\\ 
\hat{E}_{NX}&=\bigl[\hat{\Pi}_{N,A}^+-\hat{\Pi}_{N,A}^-\bigr]\!\otimes\bigl[\hat{\Pi}_{X,B}^+-\hat{\Pi}_{X,B}^-\bigr],\\ 
\hat{E}_{NN}&=\bigl[\hat{\Pi}_{N,A}^+-\hat{\Pi}_{N,A}^-\bigr]\!\otimes\bigl[\hat{\Pi}_{N,B}^+-\hat{\Pi}_{N,B}^-\bigr],
\end{align}
\end{subequations}
describe the correlations between Alice and Bob. 

First, let us study the loss-less case where $\hat{\rho}=\ket{\Psi_2}\!\bra{\Psi_2}$. Due to the assumptions $\varphi=0$ and $\zeta_A=\zeta_B$, we find $\hat{\rho}$ to be invariant when swapping the modes. In addition, the aforementioned setup to test Bell's inequality remains invariant when the observers are swapped. Thus, $\overline{E}_{XN}=\mathrm{tr}\{\hat{\rho}\hat{E}_{XN}\}=\mathrm{tr}\{\hat{\rho}\hat{E}_{NX}\}=\overline{E}_{NX}$ holds true and Eq. (\ref{eq:mean_val_Bell}) simplifies to

\begin{align}
\label{eq:mean_val_Bell_simple}
\overline{\mathcal{B}}=\overline{E}_{XX}+2\overline{E}_{XN}-\overline{E}_{NN}\leq 2.
\end{align}
In the Appendix, the expectation values of the three correlations $\overline{E}_{XX}=\mathrm{tr}\{\hat{\rho}\hat{E}_{XX}\}$, $\overline{E}_{NN}=\mathrm{tr}\{\hat{\rho}\hat{E}_{NN}\}$ and $\overline{E}_{XN}=\mathrm{tr}\{\hat{\rho}\hat{E}_{XN}\}$ are explicitly calculated. 

%
%
\begin{figure}[!ht]
\centering
\includegraphics[width=8.6cm]{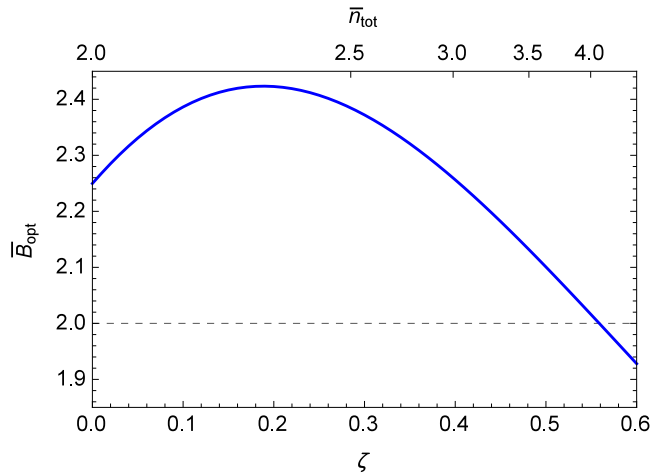}
\caption{Optimized expectation value of the Bell observable $\overline{\mathcal{B}}_\mathrm{opt}$ in dependence of the parametric gain $\zeta$ and the mean total photon number $\bar{n}_\mathrm{tot}$. The dashed horizontal line refers to $\overline{\mathcal{B}}_\mathrm{opt}=2$, indicating violation of Bell's inequality for the points lying above.}\label{fig.bell_opt_g} 
\end{figure}
%
%
%

For a given parametric gain $\zeta$, $\overline{\mathcal{B}}=\mathrm{tr}\{\hat{\rho}\hat{\mathcal{B}}\}$ is still a function of the thresholds $n_0$ and $x_0$. An optimization with respect to $n_0$ and $x_0$ for fixed $\zeta$ yields the maximal possible mean value of the Bell observable $\overline{\mathcal{B}}_\mathrm{opt}$. Fig.~\ref{fig.bell_opt_g} reports the dependence of $\overline{\mathcal{B}}_\mathrm{opt}$ on the parametric gain $\zeta$. We find violation of Bell's inequality for \textcolor{black}{$0\leq\zeta\leq0.557$}. In the case of no amplification ($\zeta=0$) we find $\overline{B}_\mathrm{opt}=2.250$. This result is in accordance with Ref.~\cite{2photonbell}, examining the case of ordinary two-photon N00N states $\ket{\psi_2}$. The maximal violation $\overline{\mathcal{B}}_\mathrm{opt}=2.423$ is obtained at \textcolor{black}{$\zeta=0.189$} for the setting $n_0=0$ and \textcolor{black}{$x_0=0.465$}. 
The optimal state contains, with a mean photon number $\bar{n}_\mathrm{tot}=2.22$, only slightly more photons than the two-photon N00N state $\ket{\psi_2}$. The maximal violation possible for two-photon states in the setup considered here is $2.46$ and has been found for an unfeasible quantum state \cite{Quintino}. We emphasize that the value $2.423$ obtained in our work is very close to this limit.
The largest value of the parametric gain for which violation occurs is $\zeta=0.557$ and corresponds to $\bar{n}_\mathrm{tot}=4.06$. All maximal values of $\overline{\mathcal{B}}_\mathrm{opt}$ have been observed for $n_0=0$, i.e., the detectors in an experiment only have to discriminate between the events \textit{photons observed} and \textit{no photons present} (threshold detectors). 

In order to understand the behavior of $\overline{\mathcal{B}}_\mathrm{opt}$ in Fig.~\ref{fig.bell_opt_g}, we analyze the three contributions to $\overline{B}_\mathrm{opt}$, namely $\overline{E}_{XX}^\mathrm{opt}$, $\overline{E}_{XN}^\mathrm{opt}$ and $\overline{E}_{NN}^\mathrm{opt}$, for increasing parametric gain $\zeta$. In the case of no amplification ($\zeta=0$), the photon number measurements with threshold $n_0=0$, performed by Alice and Bob, are perfectly anti-correlated: $\overline{E}_{NN}^\mathrm{opt}=-1$. Moreover, $\overline{E}_{XN}^\mathrm{opt}=0.561$ and $\overline{E}_{XX}^\mathrm{opt}=0.128$. 
On the one hand, we find that $|\overline{E}_{NN}^\mathrm{opt}|$ decreases as $\zeta$ grows. The reason is that the probability of Alice and Bob obtaining the same outcome ($b_A=b_B$) increases, when making the discrete variable measurement (NN). In fact, the terms $|\avg{n|\Phi_0}|^2$ with $n>0$ and $|\avg{0|\Phi_2}|^2$ grow with $\zeta$, whereas they vanish for $\zeta=0$. On the other hand, we observe that $\overline{E}_{XN}^\mathrm{opt}$ and $\overline{E}_{XX}^\mathrm{opt}$ increase with $\zeta$. This is due to the fact that with growing $\zeta$ the optimal choice of $x_0$ increases $\overline{E}_{XN}^\mathrm{opt}$ and $\overline{E}_{XX}^\mathrm{opt}$ with respect to the case of no amplification. For $\zeta<0.19$ the growth of the correlations $\overline{E}_{XN}^\mathrm{opt}$ and $\overline{E}_{XX}^\mathrm{opt}$ exceeds the drop of $|\overline{E}_{NN}^\mathrm{opt}|$. In contrast, for $\zeta>0.19$, the decrease of the anti-correlation can no longer be compensated.

Finally, we consider the violation of Bell's inequality for our scheme in the presence of losses. In general there are two types of experimental imperfection that shall be considered. The first type of loss \textcolor{black}{occurs on transmission of the state from the source to the observers and depends on the transmittance $t$ of the channels. We assume $t$ to be identical for both transmission channels.
The second type of experimental imperfections involves the detection efficiency. The efficiency of homodyne detection $\eta_X$ is usually close to 100\%, but the efficiency of quantum photon detectors $\eta_N$ in particular is significantly smaller.}

%
%
%
\begin{figure}[!ht]
\centering
\includegraphics[width=8.6cm]{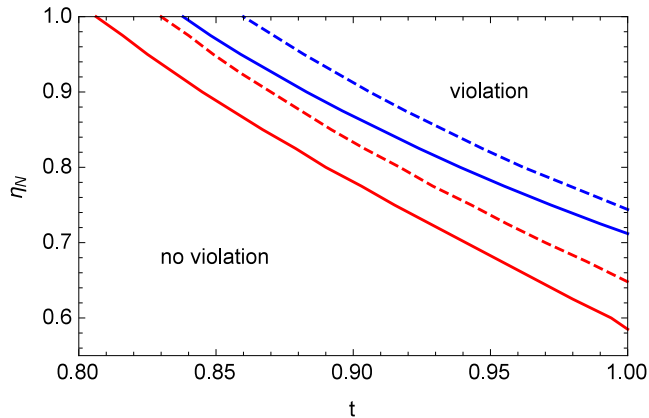}
\caption{Violation of Bell's inequality ($\overline{\mathcal{B}}_\mathrm{opt}>2$) in dependence of the transmittance $t$ and efficiency of threshold detector $\eta_N$ for different values of the parametric gain $\zeta$ and the homodyne detection efficiency $\eta_X$. \textcolor{black}{Red lines refer to an optimized gain $\zeta$}, whereas the blue lines recall the results of~\cite{2photonbell} ($\zeta=0$). For the solid lines the homodyne detection is assumed to be efficient ($\eta_X=100\%$) whereas the dashed lines correspond to a detection efficiency of $\eta_X=95\%$. Violation of Bell's inequality is attained for pairs of parameters $(t,\eta_N)$ that lie above these curves.}\label{fig.summary_lossy_belltest} 
\end{figure}
%
%
%

\textcolor{black}{The two-mode density operator
\begin{align}
\label{eq:arbitrary_density_operator}
\hat{\rho}=\ket{\Psi_2}\!\bra{\Psi_2}=\sum_{m,n=0}^\infty\sum_{m',n'=0}^\infty\rho_{mm'|nn'}\ket{m}\!\bra{n}_A\otimes\ket{m'}\!\bra{n'}_B,
\end{align}
that suffers from amplitude damping evolves into a density operator $\hat{\rho}^{(\lambda_A,\lambda_B)}$ with matrix elements
\begin{align*}
\rho_{mm'|nn'}^{(\lambda_A,\lambda_B)}&=\sum_{k=0}^\infty\sum_{k'=0}^\infty\rho_{k+m,k'+m'|k+n,k'+n'}\sqrt{B(k|\lambda_A,k+m)B(k|\lambda_A,k+n)}\sqrt{B(k'|\lambda_B,k'+m')B(k'|\lambda_B,k'+n')},
\end{align*}
where $\lambda_A$ ($\lambda_B$) is the probability to lose one photon in mode $A$ ($B$) and $B(k|p,n)={n \choose k} p^k (1-p)^{n-k}$ for $k\in\{0,\dots,n\}$ denotes the probability mass function of the binomial distribution \cite{NielsenChuangBook}. 
To study the experimental imperfections, we evaluate the three constituents of Eq.~(\ref{eq:mean_val_Bell_simple}) with 
$\hat{\rho}^{(\lambda_A,\lambda_B)}$ as 
\begin{align*}
\overline{E}_{XX}=\mathrm{tr}\{\hat{\rho}^{(1-t\eta_X,1-t\eta_X)}\hat{E}_{XX}\},~
\overline{E}_{XN}=\mathrm{tr}\{\hat{\rho}^{(1-t\eta_X,1-t\eta_N)}\hat{E}_{XN}\}\mathrm{~~and~~}
\overline{E}_{NN}=\mathrm{tr}\{\hat{\rho}^{(1-t\eta_N,1-t\eta_N)}\hat{E}_{NN}\}. 
\end{align*}
That is, we effectively assign the detection losses to the transmission channel \cite{PhysRevA.36.1955,GardinerZollerBook}. Fig.~\ref{fig.summary_lossy_belltest} summarizes the results of the analysis of experimental imperfections. It shows that the proposed scheme is robust against losses.}

Taking the homodyne detection to be efficient and assuming no transmission losses ($t=100\%$), the lowest efficiency of the threshold detector tolerable to obtain a violation of Bell's inequality is \textcolor{black}{$\eta_N^\mathrm{min}=58.5\%$}. In turn, assuming ideal detection ($\eta_N=100\%$), the highest transmission losses acceptable are \textcolor{black}{19.4\% ($t^\mathrm{min}=80.6\%$)}\footnote{\textcolor{black}{If the two-photon N00N state is not amplified at the source but instead transmitted to Alice and Bob before amplification, we find $t^\mathrm{min}=79.1\%$.}}. \textcolor{black}{The losses tolerable for a violation of Bell's inequality attained here are significantly higher than those determined for two-photon N00N states $\ket{\psi_2}$~\cite{2photonbell}. Also we find an improvement for $\eta_N^\mathrm{min}$ over the value reported for a hybrid Bell setup operating with states of the form $\ket{\psi_\mathrm{E}(\lambda)}=\ket{0}_A\ket{0}_B+\lambda\ket{\psi_2}+\mathcal{O}(\lambda^2)$, which are engineered by probabilistic amplification~\cite{Brask}. The according values of $t^\mathrm{min}$ are similar. Despite the state $\ket{P_H}$ defined in Eq. (17) of~\cite{PhysRevA.86.030101} violates the CHSH inequality for arbitrary low $\eta_N$, the minimal transmittance is given by $t^\mathrm{min}\approx92\%$. Tab.~\ref{tab:comparison} summarizes the comparison with the results obtained in~\cite{2photonbell},~\cite{Brask} and~\cite{PhysRevA.86.030101}.}
%
%
%
\begin{table}[!ht]
\begin{tabular}{lllll}\hline\hline
~ & \qquad$\ket{\Psi_2}$ & \qquad$\ket{\psi_2}$ & \qquad$\ket{\psi_\mathrm{E}(\lambda)}$ & \qquad\textcolor{black}{$\ket{P_H}$} \\
$\eta_N^\mathrm{min}$ & \qquad$58.5\%$ & \qquad$71.1\%$ & \qquad$64.8\%$ & \textcolor{black}{\qquad \;---} \\
$t^\mathrm{min}$ & \qquad$80.6\%$ & \qquad$84.0\%$ & \qquad$80.5\%$ & \textcolor{black}{\qquad$92\%$} \\\hline\hline
\end{tabular}
\caption{\label{tab:comparison}Comparing losses acceptable to violate the CHSH inequality for different quantum states. \textcolor{black}{The first column gives the results obtained here and the other columns report values found in~\cite{2photonbell},~\cite{Brask} and~\cite{PhysRevA.86.030101}, respectively.}}
\end{table}
%
%
%

Values for the efficiencies of the threshold detector and the homodyne detection, which are available with current technology, are $\eta_N=79\%$ \cite{Giustina} and $\eta_X=90\%$~\cite{Zavatta}, respectively. \textcolor{black}{Under these constraints, up to 5\% of additional transmission losses ($t\geq95\%$) are acceptable in order to still violate Bell's inequality. This shows that the Bell test proposed here is remarkably robust against experimental imperfections and potentially suitable for a loophole-free Bell test performed with state-of-the-art technology.}

\section{Possible experimental implementation of the Bell test}
\label{sec:experiment}

\begin{figure}[!ht]
\centering
\includegraphics[width=6cm]{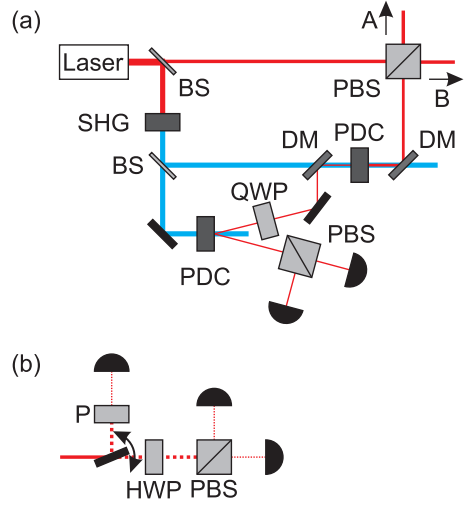}
\caption{Possible implementations of the source (a) and the measurement setup of the observers (b) using beam splitters (BS), polarizing BS (PBS), dichroic mirrors (DM), quarter-wave (QWP) and half-wave plates (HWP), polarizers (P) as well as non-linear crystals performing second harmonic generation (SHG) and parametric down-conversion (PDC).}\label{fig.possible_setup} 
\end{figure}

A possible experimental implementation of the proposed Bell test can be found in Fig.~\ref{fig.possible_setup}. It is basically a combination of the experiments~\cite{production} and~\cite{Zavatta}.
%
%
%
%
%
%
A laser beam is split on a beam splitter (BS). One part of the light beam is combined with the state $\ket{\Psi_2}$, produced in an earlier stage, on a polarizing beam splitter (PBS) and sent to the observers $A$ and $B$. The local oscillator and the signal contained in orthogonal polarization modes facilitate homodyne detection for the observer~\cite{Zavatta}. The other part of the laser light is frequency doubled by means of second harmonic generation (SHG). Part of this light is used to conditionally prepare the two-photon N00N state $(\ket{2}_A\!\ket{0}_B-\ket{0}_A\!\ket{2}_B)/\sqrt{2}$ using parametric down-conversion (PDC) and photon heralding~\cite{production}. The quarter wave plate (QWP) changes the sign such that $\ket{\psi_2}=(\ket{2}_A\!\ket{0}_B+\ket{0}_A\!\ket{2}_B)/\sqrt{2}$ is obtained. With the help of a dichroic mirror, the two-photon N00N state $\ket{\psi_2}$ is fed into an optical parametric amplifier (OPA) to produce $\ket{\Psi_2}$. On a PBS, the two polarization modes of $\ket{\Psi_2}$ are spatially separated, combined with an orthogonally polarized local oscillator field, and sent to Alice and Bob. 

With a fast switching mirror at their disposal, each observer can direct the signal randomly to a photon counting measurement or a homodyne detection scheme. By virtue of a polarizer (P), the observer can block the local oscillator and count the signal photons with a quantum photon detector. The observers homodyne detection scheme is built from a half-wave plate (HWP), a PBS, and two photo-detectors.

\section{Conclusions}
\label{sec:conclusion}

We have proposed a test of the CHSH inequality based on hybrid measurements (discrete and continuous variable) that applies to amplified two-photon N00N states. The Bell test is remarkably tolerant against experimental imperfections. In particular, the thresholds for photon detection efficiency \textcolor{black}{$\eta_N^\mathrm{min}=58.5\%$} is lower than those reported in \cite{2photonbell,Brask} for a hybrid Bell test performed with different states. \textcolor{black}{The minimal transmittance affordable $t^\mathrm{min}=80.6\%$ is smaller than the ones reported in \cite{2photonbell} and \cite{PhysRevA.86.030101} and similar to the value found in \cite{Brask}. In contrast to \cite{Brask}, the amplifier considered here works non-probabilistic.} State-of-the-art technology is \textcolor{black}{potentially} capable of performing the proposal.
Moreover, we have studied the violation of the CHSH inequality for a growing mean photon number of the amplified two-photon N00N states. For strong amplification (mean photon number $\bar{n}_\mathrm{tot}>4.06$) the suggested scheme does not show violation of Bell's inequality. The reason is that the distinguishability of the two components $\ket{\Phi_0}$ and $\ket{\Phi_2}$, cf. Eq.~(\ref{eq:an00n_state_b}), in the single-photon measurement decreases with growing parametric gain. Considering N00N states with higher photon numbers did not improve the results. In conclusion, weakly amplified two-photon N00N states \textcolor{black}{are good candidates for performing a loophole-free photonic Bell test} and for realizing several quantum information protocols. 

\section{Acknowledgment}
\textcolor{black}{We wish to thank anonymous referees for their valuable comments.} F.~T. is grateful to  M.~V. Chekhova for valuable discussions and to J. Knebel for helpful comments on the manuscript. M.~S. was supported by the EU 7FP Marie Curie Career Integration Grant No. 322150 ``QCAT'', NCN grant No. 2012/04/M/ST2/00789, FNP Homing Plus project No. HOMING PLUS/2012-5/12 and MNiSW co-financed international project No. 2586/7.PR/2012/2. The work is a part of EU project BRISQ2.

\begin{appendix}
\section{Expectation value of Bell observable}
We consider the Bell observable defined in Eq.~(\ref{eq:bell_observable}) and Eqs.~(\ref{eq:bell_correlations}) and evaluate the expectation value $\overline{\mathcal{B}}=\mathrm{tr}\{\hat{\rho}\hat{\mathcal{B}}\}$ for the arbitrary two-mode density operator from Eq.~(\ref{eq:arbitrary_density_operator}). In the following, we derive how to evaluate the correlations $\overline{E}_{XX}$, $\overline{E}_{XN}$ and $\overline{E}_{NN}$:
\begin{align*}
\overline{E}_{XX}&=\mathrm{tr}\{\hat{\rho}\bigl[\hat{\Pi}_{X,A}^+-\hat{\Pi}_{X,A}^-\bigr]\!\otimes\bigl[\hat{\Pi}_{X,B}^+-\hat{\Pi}_{X,B}^-\bigr]\}\nonumber\\
&=\sum_{m,n=0}^\infty\sum_{m',n'=0}^\infty\rho_{mm'|nn'}\left[\delta_{mn}-2\int_{-x_0}^{x_0} \rmd x\avg{n|x}\!\avg{x|m}\right]
\Biggl[\delta_{m'n'}-2\int_{-x_0}^{x_0} \rmd x\avg{n'|x}\!\avg{x|m'}\Biggr].
\end{align*}
Introducing the function 
\begin{align*}
Q_{nm}(x_0)&=\int_{-x_0}^{x_0} \rmd x\avg{n|x}\!\avg{x|m}=\sqrt{\frac{2}{\pi}}\int_{-x_0}^{x_0} \rmd x\frac{\rme^{-2x^2}}{\sqrt{2^n n!}\sqrt{2^m m!}}\mathrm{H}_n(\sqrt{2}x)\mathrm{H}_m(\sqrt{2}x),
\end{align*}
and using the fact $\tr\{\hat{\rho}\}=1$, we attain
\begin{align*}
\overline{E}_{XX}&=1+4\sum_{m,n=0}^\infty\sum_{m',n'=0}^\infty\rho_{mm'|nn'}\biggl[Q_{nm}(x_0)Q_{n'm'}(x_0)-\frac{1}{2}Q_{nm}(x_0)\delta_{m'n'}
-\frac{1}{2}\delta_{mn}Q_{n'm'}(x_0)\biggr].
\end{align*}
In order to evaluate that expression numerically, an appropriate cutoff was chosen such that $\rho_{mm'|nn'}\approx0$ for indices bigger than the cutoff. Furthermore:
\begin{align*}
\overline{E}_{XN}&=\mathrm{tr}\{\hat{\rho}\bigl[\hat{\Pi}_{X,A}^+-\hat{\Pi}_{X,A}^-\bigr]\!\otimes\bigl[\hat{\Pi}_{N,B}^+-\hat{\Pi}_{N,B}^-\bigr]\}\nonumber\\
&=2\sum_{m,n=0}^\infty\Biggl[\sum_{m'=0}^{n_0}\rho_{mm'|nm'}\bigl[\delta_{mn}-Q_{nm}(x_0)\bigr]+\!\!\sum_{m'=n_0+1}^\infty\rho_{mm'|nm'}Q_{nm}(x_0)\Biggr]-1.
\end{align*}
and finally:
\begin{align*}
\overline{E}_{NN}&=\mathrm{tr}\{\hat{\rho}\bigl[\hat{\Pi}_{N,A}^+-\hat{\Pi}_{N,A}^-\bigr]\!\otimes\bigl[\hat{\Pi}_{N,B}^+-\hat{\Pi}_{N,B}^-\bigr]\}
=1-2\sum_{m=0}^{n_0}\sum_{m'=n_0+1}^\infty\rho_{mm'|mm'}-2\sum_{m=n_0+1}^\infty\sum_{m'=0}^{n_0}\rho_{mm'|mm'}.
\end{align*}

\end{appendix}

\bibliography{bibliography}

\begin{thebibliography}{41}%
\makeatletter
\providecommand \@ifxundefined [1]{%
 \@ifx{#1\undefined}
}%
\providecommand \@ifnum [1]{%
 \ifnum #1\expandafter \@firstoftwo
 \else \expandafter \@secondoftwo
 \fi
}%
\providecommand \@ifx [1]{%
 \ifx #1\expandafter \@firstoftwo
 \else \expandafter \@secondoftwo
 \fi
}%
\providecommand \natexlab [1]{#1}%
\providecommand \enquote  [1]{``#1''}%
\providecommand \bibnamefont  [1]{#1}%
\providecommand \bibfnamefont [1]{#1}%
\providecommand \citenamefont [1]{#1}%
\providecommand \href@noop [0]{\@secondoftwo}%
\providecommand \href [0]{\begingroup \@sanitize@url \@href}%
\providecommand \@href[1]{\@@startlink{#1}\@@href}%
\providecommand \@@href[1]{\endgroup#1\@@endlink}%
\providecommand \@sanitize@url [0]{\catcode `\\12\catcode `\$12\catcode
  `\&12\catcode `\#12\catcode `\^12\catcode `\_12\catcode `\%12\relax}%
\providecommand \@@startlink[1]{}%
\providecommand \@@endlink[0]{}%
\providecommand \url  [0]{\begingroup\@sanitize@url \@url }%
\providecommand \@url [1]{\endgroup\@href {#1}{\urlprefix }}%
\providecommand \urlprefix  [0]{URL }%
\providecommand \Eprint [0]{\href }%
\providecommand \doibase [0]{http://dx.doi.org/}%
\providecommand \selectlanguage [0]{\@gobble}%
\providecommand \bibinfo  [0]{\@secondoftwo}%
\providecommand \bibfield  [0]{\@secondoftwo}%
\providecommand \translation [1]{[#1]}%
\providecommand \BibitemOpen [0]{}%
\providecommand \bibitemStop [0]{}%
\providecommand \bibitemNoStop [0]{.\EOS\space}%
\providecommand \EOS [0]{\spacefactor3000\relax}%
\providecommand \BibitemShut  [1]{\csname bibitem#1\endcsname}%
\let\auto@bib@innerbib\@empty
\bibitem [{\citenamefont {Bell}(1964)}]{Bell}%
  \BibitemOpen
  \bibfield  {author} {\bibinfo {author} {\bibfnamefont {J.}~\bibnamefont
  {Bell}},\ }\href@noop {} {\bibfield  {journal} {\bibinfo  {journal}
  {Physics}\ }\textbf {\bibinfo {volume} {1}},\ \bibinfo {pages} {195}
  (\bibinfo {year} {1964})}\BibitemShut {NoStop}%
\bibitem [{\citenamefont {Einstein}\ \emph {et~al.}(1935)\citenamefont
  {Einstein}, \citenamefont {Podolsky},\ and\ \citenamefont {Rosen}}]{EPR}%
  \BibitemOpen
  \bibfield  {author} {\bibinfo {author} {\bibfnamefont {A.}~\bibnamefont
  {Einstein}}, \bibinfo {author} {\bibfnamefont {B.}~\bibnamefont {Podolsky}},
  \ and\ \bibinfo {author} {\bibfnamefont {N.}~\bibnamefont {Rosen}},\ }\href
  {\doibase 10.1103/PhysRev.47.777} {\bibfield  {journal} {\bibinfo  {journal}
  {Phys. Rev.}\ }\textbf {\bibinfo {volume} {47}},\ \bibinfo {pages} {777}
  (\bibinfo {year} {1935})}\BibitemShut {NoStop}%
\bibitem [{\citenamefont {Brukner}\ and\ \citenamefont {\ifmmode
  {\dot{Z}}\else~{\.{Z}}\fi{}kowski}(2012)}]{Brukner}%
  \BibitemOpen
  \bibfield  {author} {\bibinfo {author} {\bibfnamefont {{\v{C}}.}~\bibnamefont
  {Brukner}}\ and\ \bibinfo {author} {\bibfnamefont {M.}~\bibnamefont {\ifmmode
  {\dot{Z}}\else~{\.{Z}}\fi{}kowski}},\ }in\ \href@noop {} {\emph {\bibinfo
  {booktitle} {Handbook of Natural Computing}}},\ \bibinfo {editor} {edited by\
  \bibinfo {editor} {\bibfnamefont {G.}~\bibnamefont {Rozenberg}}, \bibinfo
  {editor} {\bibfnamefont {T.}~\bibnamefont {B\"ack}}, \ and\ \bibinfo {editor}
  {\bibfnamefont {J.~N.}\ \bibnamefont {Kok}}}\ (\bibinfo  {publisher}
  {Springer Verlag},\ \bibinfo {year} {2012})\ \bibinfo {type}
  {Manual}~\bibinfo {chapter} {42}, pp.\ \bibinfo {pages}
  {1413--1450}\BibitemShut {NoStop}%
\bibitem [{\citenamefont {Ac\'in}\ \emph {et~al.}(2007)\citenamefont {Ac\'in},
  \citenamefont {Brunner}, \citenamefont {Gisin}, \citenamefont {Massar},
  \citenamefont {Pironio},\ and\ \citenamefont {Scarani}}]{Acin2007}%
  \BibitemOpen
  \bibfield  {author} {\bibinfo {author} {\bibfnamefont {A.}~\bibnamefont
  {Ac\'in}}, \bibinfo {author} {\bibfnamefont {N.}~\bibnamefont {Brunner}},
  \bibinfo {author} {\bibfnamefont {N.}~\bibnamefont {Gisin}}, \bibinfo
  {author} {\bibfnamefont {S.}~\bibnamefont {Massar}}, \bibinfo {author}
  {\bibfnamefont {S.}~\bibnamefont {Pironio}}, \ and\ \bibinfo {author}
  {\bibfnamefont {V.}~\bibnamefont {Scarani}},\ }\href {\doibase
  10.1103/PhysRevLett.98.230501} {\bibfield  {journal} {\bibinfo  {journal}
  {Phys. Rev. Lett.}\ }\textbf {\bibinfo {volume} {98}},\ \bibinfo {pages}
  {230501} (\bibinfo {year} {2007})}\BibitemShut {NoStop}%
\bibitem [{\citenamefont {Scarani}\ and\ \citenamefont
  {Gisin}(2001)}]{Scarani}%
  \BibitemOpen
  \bibfield  {author} {\bibinfo {author} {\bibfnamefont {V.}~\bibnamefont
  {Scarani}}\ and\ \bibinfo {author} {\bibfnamefont {N.}~\bibnamefont
  {Gisin}},\ }\href {\doibase 10.1103/PhysRevLett.87.117901} {\bibfield
  {journal} {\bibinfo  {journal} {Phys. Rev. Lett.}\ }\textbf {\bibinfo
  {volume} {87}},\ \bibinfo {pages} {117901} (\bibinfo {year}
  {2001})}\BibitemShut {NoStop}%
\bibitem [{\citenamefont {Pironio}\ \emph {et~al.}(2010)\citenamefont
  {Pironio}, \citenamefont {Ac\'in}, \citenamefont {Massar}, \citenamefont
  {Boye de~la Giroday}, \citenamefont {Matsukevich}, \citenamefont {Maunz},
  \citenamefont {Olmschenk}, \citenamefont {Hayes}, \citenamefont {Luo},
  \citenamefont {Manning},\ and\ \citenamefont {Monroe}}]{Pironio2010}%
  \BibitemOpen
  \bibfield  {author} {\bibinfo {author} {\bibfnamefont {S.}~\bibnamefont
  {Pironio}}, \bibinfo {author} {\bibfnamefont {A.}~\bibnamefont {Ac\'in}},
  \bibinfo {author} {\bibfnamefont {S.}~\bibnamefont {Massar}}, \bibinfo
  {author} {\bibfnamefont {A.}~\bibnamefont {Boye de~la Giroday}}, \bibinfo
  {author} {\bibfnamefont {D.~N.}\ \bibnamefont {Matsukevich}}, \bibinfo
  {author} {\bibfnamefont {P.}~\bibnamefont {Maunz}}, \bibinfo {author}
  {\bibfnamefont {S.}~\bibnamefont {Olmschenk}}, \bibinfo {author}
  {\bibfnamefont {D.}~\bibnamefont {Hayes}}, \bibinfo {author} {\bibfnamefont
  {L.}~\bibnamefont {Luo}}, \bibinfo {author} {\bibfnamefont {T.~A.}\
  \bibnamefont {Manning}}, \ and\ \bibinfo {author} {\bibfnamefont
  {C.}~\bibnamefont {Monroe}},\ }\href {\doibase 10.1038/nature09008}
  {\bibfield  {journal} {\bibinfo  {journal} {Nature}\ }\textbf {\bibinfo
  {volume} {464}},\ \bibinfo {pages} {1021} (\bibinfo {year}
  {2010})}\BibitemShut {NoStop}%
\bibitem [{\citenamefont {Brukner}\ \emph {et~al.}(2004)\citenamefont
  {Brukner}, \citenamefont {\ifmmode {\dot{Z}}\else~{\.{Z}}\fi{}ukowski},
  \citenamefont {Pan},\ and\ \citenamefont {Zeilinger}}]{Zukowski}%
  \BibitemOpen
  \bibfield  {author} {\bibinfo {author} {\bibfnamefont {{\v{C}}.}~\bibnamefont
  {Brukner}}, \bibinfo {author} {\bibfnamefont {M.}~\bibnamefont {\ifmmode
  {\dot{Z}}\else~{\.{Z}}\fi{}ukowski}}, \bibinfo {author} {\bibfnamefont
  {J.-W.}\ \bibnamefont {Pan}}, \ and\ \bibinfo {author} {\bibfnamefont
  {A.}~\bibnamefont {Zeilinger}},\ }\href {\doibase
  10.1103/PhysRevLett.92.127901} {\bibfield  {journal} {\bibinfo  {journal}
  {Phys. Rev. Lett.}\ }\textbf {\bibinfo {volume} {92}},\ \bibinfo {pages}
  {127901} (\bibinfo {year} {2004})}\BibitemShut {NoStop}%
\bibitem [{\citenamefont {Aspect}\ \emph {et~al.}(1981)\citenamefont {Aspect},
  \citenamefont {Grangier},\ and\ \citenamefont {Roger}}]{Aspect}%
  \BibitemOpen
  \bibfield  {author} {\bibinfo {author} {\bibfnamefont {A.}~\bibnamefont
  {Aspect}}, \bibinfo {author} {\bibfnamefont {P.}~\bibnamefont {Grangier}}, \
  and\ \bibinfo {author} {\bibfnamefont {G.}~\bibnamefont {Roger}},\ }\href
  {\doibase 10.1103/PhysRevLett.47.460} {\bibfield  {journal} {\bibinfo
  {journal} {Phys. Rev. Lett.}\ }\textbf {\bibinfo {volume} {47}},\ \bibinfo
  {pages} {460} (\bibinfo {year} {1981})}\BibitemShut {NoStop}%
\bibitem [{\citenamefont {Hofmann}\ \emph {et~al.}(2012)\citenamefont
  {Hofmann}, \citenamefont {Krug}, \citenamefont {Ortege}, \citenamefont
  {G\'erard}, \citenamefont {Weber}, \citenamefont {Rosenfeld},\ and\
  \citenamefont {Weinfurter}}]{Hofmann}%
  \BibitemOpen
  \bibfield  {author} {\bibinfo {author} {\bibfnamefont {J.}~\bibnamefont
  {Hofmann}}, \bibinfo {author} {\bibfnamefont {M.}~\bibnamefont {Krug}},
  \bibinfo {author} {\bibfnamefont {l.~N.}\ \bibnamefont {Ortege}}, \bibinfo
  {author} {\bibfnamefont {L.}~\bibnamefont {G\'erard}}, \bibinfo {author}
  {\bibfnamefont {M.}~\bibnamefont {Weber}}, \bibinfo {author} {\bibfnamefont
  {W.}~\bibnamefont {Rosenfeld}}, \ and\ \bibinfo {author} {\bibfnamefont
  {H.}~\bibnamefont {Weinfurter}},\ }\href@noop {} {\bibfield  {journal}
  {\bibinfo  {journal} {Science}\ }\textbf {\bibinfo {volume} {337}},\ \bibinfo
  {pages} {72} (\bibinfo {year} {2012})}\BibitemShut {NoStop}%
\bibitem [{\citenamefont {Rowe}\ \emph {et~al.}(2001)\citenamefont {Rowe},
  \citenamefont {Kielpinski}, \citenamefont {Meyer}, \citenamefont {Sackett},
  \citenamefont {Itano}, \citenamefont {Monroe},\ and\ \citenamefont
  {Wineland}}]{Rowe}%
  \BibitemOpen
  \bibfield  {author} {\bibinfo {author} {\bibfnamefont {M.~A.}\ \bibnamefont
  {Rowe}}, \bibinfo {author} {\bibfnamefont {D.~V.}\ \bibnamefont
  {Kielpinski}}, \bibinfo {author} {\bibfnamefont {V.}~\bibnamefont {Meyer}},
  \bibinfo {author} {\bibfnamefont {C.~A.}\ \bibnamefont {Sackett}}, \bibinfo
  {author} {\bibfnamefont {W.~M.}\ \bibnamefont {Itano}}, \bibinfo {author}
  {\bibfnamefont {C.}~\bibnamefont {Monroe}}, \ and\ \bibinfo {author}
  {\bibfnamefont {D.~J.}\ \bibnamefont {Wineland}},\ }\href@noop {} {\bibfield
  {journal} {\bibinfo  {journal} {Nature}\ }\textbf {\bibinfo {volume} {409}},\
  \bibinfo {pages} {791} (\bibinfo {year} {2001})}\BibitemShut {NoStop}%
\bibitem [{\citenamefont {Ansmann}\ \emph {et~al.}(2009)\citenamefont
  {Ansmann}, \citenamefont {Wang}, \citenamefont {Bialczak}, \citenamefont
  {Hofheinz}, \citenamefont {Lucero}, \citenamefont {Neeley}, \citenamefont
  {O'Connell}, \citenamefont {Sank}, \citenamefont {Weides}, \citenamefont
  {Wenner}, \citenamefont {Cleland},\ and\ \citenamefont {Martinis}}]{Ansmann}%
  \BibitemOpen
  \bibfield  {author} {\bibinfo {author} {\bibfnamefont {M.}~\bibnamefont
  {Ansmann}}, \bibinfo {author} {\bibfnamefont {H.}~\bibnamefont {Wang}},
  \bibinfo {author} {\bibfnamefont {R.~C.}\ \bibnamefont {Bialczak}}, \bibinfo
  {author} {\bibfnamefont {M.}~\bibnamefont {Hofheinz}}, \bibinfo {author}
  {\bibfnamefont {E.}~\bibnamefont {Lucero}}, \bibinfo {author} {\bibfnamefont
  {M.}~\bibnamefont {Neeley}}, \bibinfo {author} {\bibfnamefont {A.~D.}\
  \bibnamefont {O'Connell}}, \bibinfo {author} {\bibfnamefont {D.}~\bibnamefont
  {Sank}}, \bibinfo {author} {\bibfnamefont {M.}~\bibnamefont {Weides}},
  \bibinfo {author} {\bibfnamefont {J.}~\bibnamefont {Wenner}}, \bibinfo
  {author} {\bibfnamefont {A.~N.}\ \bibnamefont {Cleland}}, \ and\ \bibinfo
  {author} {\bibfnamefont {J.}~\bibnamefont {Martinis}},\ }\href {\doibase
  10.1038/nature08363} {\bibfield  {journal} {\bibinfo  {journal} {Nature}\
  }\textbf {\bibinfo {volume} {461}},\ \bibinfo {pages} {504} (\bibinfo {year}
  {2009})}\BibitemShut {NoStop}%
\bibitem [{\citenamefont {Giustina}\ \emph {et~al.}(2013)\citenamefont
  {Giustina}, \citenamefont {Mech}, \citenamefont {Ramelow}, \citenamefont
  {Wittmann}, \citenamefont {Kofler}, \citenamefont {Beyer}, \citenamefont
  {Lita}, \citenamefont {Calkins}, \citenamefont {Gerrits}, \citenamefont
  {Nam}, \citenamefont {Ursin},\ and\ \citenamefont {Zeilinger}}]{Giustina}%
  \BibitemOpen
  \bibfield  {author} {\bibinfo {author} {\bibfnamefont {M.}~\bibnamefont
  {Giustina}}, \bibinfo {author} {\bibfnamefont {A.}~\bibnamefont {Mech}},
  \bibinfo {author} {\bibfnamefont {S.}~\bibnamefont {Ramelow}}, \bibinfo
  {author} {\bibfnamefont {B.}~\bibnamefont {Wittmann}}, \bibinfo {author}
  {\bibfnamefont {J.}~\bibnamefont {Kofler}}, \bibinfo {author} {\bibfnamefont
  {J.}~\bibnamefont {Beyer}}, \bibinfo {author} {\bibfnamefont
  {A.}~\bibnamefont {Lita}}, \bibinfo {author} {\bibfnamefont {B.}~\bibnamefont
  {Calkins}}, \bibinfo {author} {\bibfnamefont {T.}~\bibnamefont {Gerrits}},
  \bibinfo {author} {\bibfnamefont {S.~W.}\ \bibnamefont {Nam}}, \bibinfo
  {author} {\bibfnamefont {R.}~\bibnamefont {Ursin}}, \ and\ \bibinfo {author}
  {\bibfnamefont {A.}~\bibnamefont {Zeilinger}},\ }\href {\doibase
  10.1038/nature12012} {\bibfield  {journal} {\bibinfo  {journal} {Nature}\
  }\textbf {\bibinfo {volume} {497}},\ \bibinfo {pages} {227} (\bibinfo {year}
  {2013})}\BibitemShut {NoStop}%
\bibitem [{\citenamefont {Christensen}\ \emph {et~al.}(2013)\citenamefont
  {Christensen}, \citenamefont {McCusker}, \citenamefont {Altepeter},
  \citenamefont {Calkins}, \citenamefont {Gerrits}, \citenamefont {Lita},
  \citenamefont {Miller}, \citenamefont {Shalm}, \citenamefont {Zhang},
  \citenamefont {Nam}, \citenamefont {Brunner}, \citenamefont {Lim},
  \citenamefont {Gisin},\ and\ \citenamefont {Kwiat}}]{Christensen}%
  \BibitemOpen
  \bibfield  {author} {\bibinfo {author} {\bibfnamefont {B.~G.}\ \bibnamefont
  {Christensen}}, \bibinfo {author} {\bibfnamefont {K.~T.}\ \bibnamefont
  {McCusker}}, \bibinfo {author} {\bibfnamefont {J.~B.}\ \bibnamefont
  {Altepeter}}, \bibinfo {author} {\bibfnamefont {B.}~\bibnamefont {Calkins}},
  \bibinfo {author} {\bibfnamefont {T.}~\bibnamefont {Gerrits}}, \bibinfo
  {author} {\bibfnamefont {A.~E.}\ \bibnamefont {Lita}}, \bibinfo {author}
  {\bibfnamefont {A.}~\bibnamefont {Miller}}, \bibinfo {author} {\bibfnamefont
  {L.~K.}\ \bibnamefont {Shalm}}, \bibinfo {author} {\bibfnamefont
  {Y.}~\bibnamefont {Zhang}}, \bibinfo {author} {\bibfnamefont {S.~W.}\
  \bibnamefont {Nam}}, \bibinfo {author} {\bibfnamefont {N.}~\bibnamefont
  {Brunner}}, \bibinfo {author} {\bibfnamefont {C.~C.~W.}\ \bibnamefont {Lim}},
  \bibinfo {author} {\bibfnamefont {N.}~\bibnamefont {Gisin}}, \ and\ \bibinfo
  {author} {\bibfnamefont {P.~G.}\ \bibnamefont {Kwiat}},\ }\href {\doibase
  10.1103/PhysRevLett.111.130406} {\bibfield  {journal} {\bibinfo  {journal}
  {Phys. Rev. Lett.}\ }\textbf {\bibinfo {volume} {111}},\ \bibinfo {pages}
  {130406} (\bibinfo {year} {2013})}\BibitemShut {NoStop}%
\bibitem [{\citenamefont {Lita}\ \emph {et~al.}(2008)\citenamefont {Lita},
  \citenamefont {Miller},\ and\ \citenamefont {Nam}}]{detector1}%
  \BibitemOpen
  \bibfield  {author} {\bibinfo {author} {\bibfnamefont {A.~E.}\ \bibnamefont
  {Lita}}, \bibinfo {author} {\bibfnamefont {A.~J.}\ \bibnamefont {Miller}}, \
  and\ \bibinfo {author} {\bibfnamefont {S.~W.}\ \bibnamefont {Nam}},\ }\href
  {\doibase 10.1364/OE.16.003032} {\bibfield  {journal} {\bibinfo  {journal}
  {Opt. Express}\ }\textbf {\bibinfo {volume} {16}},\ \bibinfo {pages} {3032}
  (\bibinfo {year} {2008})}\BibitemShut {NoStop}%
\bibitem [{\citenamefont {Gerrits}\ \emph {et~al.}(2012)\citenamefont
  {Gerrits}, \citenamefont {Calkins}, \citenamefont {Tomlin}, \citenamefont
  {Lita}, \citenamefont {Migdall}, \citenamefont {Mirin},\ and\ \citenamefont
  {Nam}}]{detector2}%
  \BibitemOpen
  \bibfield  {author} {\bibinfo {author} {\bibfnamefont {T.}~\bibnamefont
  {Gerrits}}, \bibinfo {author} {\bibfnamefont {B.}~\bibnamefont {Calkins}},
  \bibinfo {author} {\bibfnamefont {N.}~\bibnamefont {Tomlin}}, \bibinfo
  {author} {\bibfnamefont {A.~E.}\ \bibnamefont {Lita}}, \bibinfo {author}
  {\bibfnamefont {A.}~\bibnamefont {Migdall}}, \bibinfo {author} {\bibfnamefont
  {R.}~\bibnamefont {Mirin}}, \ and\ \bibinfo {author} {\bibfnamefont {S.~W.}\
  \bibnamefont {Nam}},\ }\href {\doibase 10.1364/OE.20.023798} {\bibfield
  {journal} {\bibinfo  {journal} {Opt. Express}\ }\textbf {\bibinfo {volume}
  {20}},\ \bibinfo {pages} {23798} (\bibinfo {year} {2012})}\BibitemShut
  {NoStop}%
\bibitem [{\citenamefont {Calkins}\ \emph {et~al.}(2013)\citenamefont
  {Calkins}, \citenamefont {Mennea}, \citenamefont {Lita}, \citenamefont
  {Metcalf}, \citenamefont {Kolthammer}, \citenamefont {Lamas-Linares},
  \citenamefont {Spring}, \citenamefont {Humphreys}, \citenamefont {Mirin},
  \citenamefont {Gates}, \citenamefont {Smith}, \citenamefont {Walmsley},
  \citenamefont {Gerrits},\ and\ \citenamefont {Nam}}]{detector3}%
  \BibitemOpen
  \bibfield  {author} {\bibinfo {author} {\bibfnamefont {B.}~\bibnamefont
  {Calkins}}, \bibinfo {author} {\bibfnamefont {P.~L.}\ \bibnamefont {Mennea}},
  \bibinfo {author} {\bibfnamefont {A.~E.}\ \bibnamefont {Lita}}, \bibinfo
  {author} {\bibfnamefont {B.~J.}\ \bibnamefont {Metcalf}}, \bibinfo {author}
  {\bibfnamefont {W.~S.}\ \bibnamefont {Kolthammer}}, \bibinfo {author}
  {\bibfnamefont {A.}~\bibnamefont {Lamas-Linares}}, \bibinfo {author}
  {\bibfnamefont {J.~B.}\ \bibnamefont {Spring}}, \bibinfo {author}
  {\bibfnamefont {P.~C.}\ \bibnamefont {Humphreys}}, \bibinfo {author}
  {\bibfnamefont {R.~P.}\ \bibnamefont {Mirin}}, \bibinfo {author}
  {\bibfnamefont {J.~C.}\ \bibnamefont {Gates}}, \bibinfo {author}
  {\bibfnamefont {P.~G.~R.}\ \bibnamefont {Smith}}, \bibinfo {author}
  {\bibfnamefont {I.~A.}\ \bibnamefont {Walmsley}}, \bibinfo {author}
  {\bibfnamefont {T.}~\bibnamefont {Gerrits}}, \ and\ \bibinfo {author}
  {\bibfnamefont {S.~W.}\ \bibnamefont {Nam}},\ }\href {\doibase
  10.1364/OE.21.022657} {\bibfield  {journal} {\bibinfo  {journal} {Opt.
  Express}\ }\textbf {\bibinfo {volume} {21}},\ \bibinfo {pages} {22657}
  (\bibinfo {year} {2013})}\BibitemShut {NoStop}%
\bibitem [{\citenamefont {Gilchrist}\ \emph {et~al.}(1998)\citenamefont
  {Gilchrist}, \citenamefont {Deuar},\ and\ \citenamefont {Reid}}]{Gilchrist}%
  \BibitemOpen
  \bibfield  {author} {\bibinfo {author} {\bibfnamefont {A.}~\bibnamefont
  {Gilchrist}}, \bibinfo {author} {\bibfnamefont {P.}~\bibnamefont {Deuar}}, \
  and\ \bibinfo {author} {\bibfnamefont {M.~D.}\ \bibnamefont {Reid}},\ }\href
  {\doibase 10.1103/PhysRevLett.80.3169} {\bibfield  {journal} {\bibinfo
  {journal} {Phys. Rev. Lett.}\ }\textbf {\bibinfo {volume} {80}},\ \bibinfo
  {pages} {3169} (\bibinfo {year} {1998})}\BibitemShut {NoStop}%
\bibitem [{\citenamefont {Cavalcanti}\ \emph {et~al.}(2011)\citenamefont
  {Cavalcanti}, \citenamefont {Brunner}, \citenamefont {Skrzypczyk},
  \citenamefont {Salles},\ and\ \citenamefont {Scarani}}]{2photonbell}%
  \BibitemOpen
  \bibfield  {author} {\bibinfo {author} {\bibfnamefont {D.}~\bibnamefont
  {Cavalcanti}}, \bibinfo {author} {\bibfnamefont {N.}~\bibnamefont {Brunner}},
  \bibinfo {author} {\bibfnamefont {P.}~\bibnamefont {Skrzypczyk}}, \bibinfo
  {author} {\bibfnamefont {A.}~\bibnamefont {Salles}}, \ and\ \bibinfo {author}
  {\bibfnamefont {V.}~\bibnamefont {Scarani}},\ }\href {\doibase
  10.1103/PhysRevA.84.022105} {\bibfield  {journal} {\bibinfo  {journal} {Phys.
  Rev. A}\ }\textbf {\bibinfo {volume} {84}},\ \bibinfo {pages} {022105}
  (\bibinfo {year} {2011})}\BibitemShut {NoStop}%
\bibitem [{\citenamefont {Brask}\ \emph {et~al.}(2012)\citenamefont {Brask},
  \citenamefont {Brunner}, \citenamefont {Cavalcanti},\ and\ \citenamefont
  {Leverrier}}]{Brask}%
  \BibitemOpen
  \bibfield  {author} {\bibinfo {author} {\bibfnamefont {J.~B.}\ \bibnamefont
  {Brask}}, \bibinfo {author} {\bibfnamefont {N.}~\bibnamefont {Brunner}},
  \bibinfo {author} {\bibfnamefont {D.}~\bibnamefont {Cavalcanti}}, \ and\
  \bibinfo {author} {\bibfnamefont {A.}~\bibnamefont {Leverrier}},\ }\href
  {\doibase 10.1103/PhysRevA.85.042116} {\bibfield  {journal} {\bibinfo
  {journal} {Phys. Rev. A}\ }\textbf {\bibinfo {volume} {85}},\ \bibinfo
  {pages} {042116} (\bibinfo {year} {2012})}\BibitemShut {NoStop}%
\bibitem [{\citenamefont {Quintino}\ \emph {et~al.}(2012)\citenamefont
  {Quintino}, \citenamefont {Ara\'ujo}, \citenamefont {Cavalcanti},
  \citenamefont {Santos},\ and\ \citenamefont {Cunha}}]{Quintino}%
  \BibitemOpen
  \bibfield  {author} {\bibinfo {author} {\bibfnamefont {M.~T.}\ \bibnamefont
  {Quintino}}, \bibinfo {author} {\bibfnamefont {M.}~\bibnamefont {Ara\'ujo}},
  \bibinfo {author} {\bibfnamefont {D.}~\bibnamefont {Cavalcanti}}, \bibinfo
  {author} {\bibfnamefont {M.~F.}\ \bibnamefont {Santos}}, \ and\ \bibinfo
  {author} {\bibfnamefont {M.~T.}\ \bibnamefont {Cunha}},\ }\href
  {http://stacks.iop.org/1751-8121/45/i=21/a=215308} {\bibfield  {journal}
  {\bibinfo  {journal} {J. Phys. A: Math. Theor.}\ }\textbf {\bibinfo {volume}
  {45}},\ \bibinfo {pages} {215308} (\bibinfo {year} {2012})}\BibitemShut
  {NoStop}%
\bibitem [{\citenamefont {Sangouard}\ \emph {et~al.}(2011)\citenamefont
  {Sangouard}, \citenamefont {Bancal}, \citenamefont {Gisin}, \citenamefont
  {Rosenfeld}, \citenamefont {Sekatski}, \citenamefont {Weber},\ and\
  \citenamefont {Weinfurter}}]{PhysRevA.84.052122}%
  \BibitemOpen
  \bibfield  {author} {\bibinfo {author} {\bibfnamefont {N.}~\bibnamefont
  {Sangouard}}, \bibinfo {author} {\bibfnamefont {J.-D.}\ \bibnamefont
  {Bancal}}, \bibinfo {author} {\bibfnamefont {N.}~\bibnamefont {Gisin}},
  \bibinfo {author} {\bibfnamefont {W.}~\bibnamefont {Rosenfeld}}, \bibinfo
  {author} {\bibfnamefont {P.}~\bibnamefont {Sekatski}}, \bibinfo {author}
  {\bibfnamefont {M.}~\bibnamefont {Weber}}, \ and\ \bibinfo {author}
  {\bibfnamefont {H.}~\bibnamefont {Weinfurter}},\ }\href {\doibase
  10.1103/PhysRevA.84.052122} {\bibfield  {journal} {\bibinfo  {journal} {Phys.
  Rev. A}\ }\textbf {\bibinfo {volume} {84}},\ \bibinfo {pages} {052122}
  (\bibinfo {year} {2011})}\BibitemShut {NoStop}%
\bibitem [{\citenamefont {Teo}\ \emph {et~al.}(2013)\citenamefont {Teo},
  \citenamefont {Araújo}, \citenamefont {Quintino}, \citenamefont {Minář},
  \citenamefont {Cavalcanti}, \citenamefont {Scarani}, \citenamefont
  {Terra~Cunha},\ and\ \citenamefont {França~Santos}}]{Teo}%
  \BibitemOpen
  \bibfield  {author} {\bibinfo {author} {\bibfnamefont {C.}~\bibnamefont
  {Teo}}, \bibinfo {author} {\bibfnamefont {M.}~\bibnamefont {Araújo}},
  \bibinfo {author} {\bibfnamefont {M.~T.}\ \bibnamefont {Quintino}}, \bibinfo
  {author} {\bibfnamefont {J.}~\bibnamefont {Minář}}, \bibinfo {author}
  {\bibfnamefont {D.}~\bibnamefont {Cavalcanti}}, \bibinfo {author}
  {\bibfnamefont {V.}~\bibnamefont {Scarani}}, \bibinfo {author} {\bibfnamefont
  {M.}~\bibnamefont {Terra~Cunha}}, \ and\ \bibinfo {author} {\bibfnamefont
  {M.}~\bibnamefont {França~Santos}},\ }\href {\doibase 10.1038/ncomms3104}
  {\bibfield  {journal} {\bibinfo  {journal} {Nat. Commun.}\ }\textbf {\bibinfo
  {volume} {4}},\ \bibinfo {pages} {2104} (\bibinfo {year} {2013})}\BibitemShut
  {NoStop}%
\bibitem [{\citenamefont {Ara\'ujo}\ \emph {et~al.}(2012)\citenamefont
  {Ara\'ujo}, \citenamefont {Quintino}, \citenamefont {Cavalcanti},
  \citenamefont {Santos}, \citenamefont {Cabello},\ and\ \citenamefont
  {Cunha}}]{PhysRevA.86.030101}%
  \BibitemOpen
  \bibfield  {author} {\bibinfo {author} {\bibfnamefont {M.}~\bibnamefont
  {Ara\'ujo}}, \bibinfo {author} {\bibfnamefont {M.}~\bibnamefont {Quintino}},
  \bibinfo {author} {\bibfnamefont {D.}~\bibnamefont {Cavalcanti}}, \bibinfo
  {author} {\bibfnamefont {M.}~\bibnamefont {Santos}}, \bibinfo {author}
  {\bibfnamefont {A.}~\bibnamefont {Cabello}}, \ and\ \bibinfo {author}
  {\bibfnamefont {M.}~\bibnamefont {Cunha}},\ }\href {\doibase
  10.1103/PhysRevA.86.030101} {\bibfield  {journal} {\bibinfo  {journal} {Phys.
  Rev. A}\ }\textbf {\bibinfo {volume} {86}},\ \bibinfo {pages} {030101}
  (\bibinfo {year} {2012})}\BibitemShut {NoStop}%
\bibitem [{\citenamefont {Vitelli}\ \emph {et~al.}(2010)\citenamefont
  {Vitelli}, \citenamefont {Spagnolo}, \citenamefont {Toffoli}, \citenamefont
  {Sciarrino},\ and\ \citenamefont {De~Martini}}]{Vitelli}%
  \BibitemOpen
  \bibfield  {author} {\bibinfo {author} {\bibfnamefont {C.}~\bibnamefont
  {Vitelli}}, \bibinfo {author} {\bibfnamefont {N.}~\bibnamefont {Spagnolo}},
  \bibinfo {author} {\bibfnamefont {L.}~\bibnamefont {Toffoli}}, \bibinfo
  {author} {\bibfnamefont {F.}~\bibnamefont {Sciarrino}}, \ and\ \bibinfo
  {author} {\bibfnamefont {F.}~\bibnamefont {De~Martini}},\ }\href {\doibase
  10.1103/PhysRevA.81.032123} {\bibfield  {journal} {\bibinfo  {journal} {Phys.
  Rev. A}\ }\textbf {\bibinfo {volume} {81}},\ \bibinfo {pages} {032123}
  (\bibinfo {year} {2010})}\BibitemShut {NoStop}%
\bibitem [{\citenamefont {Stobi\ifmmode~\acute{n}\else \'{n}\fi{}ska}\ \emph
  {et~al.}(2011)\citenamefont {Stobi\ifmmode~\acute{n}\else \'{n}\fi{}ska},
  \citenamefont {Sekatski}, \citenamefont {Buraczewski}, \citenamefont
  {Gisin},\ and\ \citenamefont {Leuchs}}]{Stobinska2011}%
  \BibitemOpen
  \bibfield  {author} {\bibinfo {author} {\bibfnamefont {M.}~\bibnamefont
  {Stobi\ifmmode~\acute{n}\else \'{n}\fi{}ska}}, \bibinfo {author}
  {\bibfnamefont {P.}~\bibnamefont {Sekatski}}, \bibinfo {author}
  {\bibfnamefont {A.}~\bibnamefont {Buraczewski}}, \bibinfo {author}
  {\bibfnamefont {N.}~\bibnamefont {Gisin}}, \ and\ \bibinfo {author}
  {\bibfnamefont {G.}~\bibnamefont {Leuchs}},\ }\href {\doibase
  10.1103/PhysRevA.84.034104} {\bibfield  {journal} {\bibinfo  {journal} {Phys.
  Rev. A}\ }\textbf {\bibinfo {volume} {84}},\ \bibinfo {pages} {034104}
  (\bibinfo {year} {2011})}\BibitemShut {NoStop}%
\bibitem [{\citenamefont {Sekatski}\ \emph {et~al.}(2012)\citenamefont
  {Sekatski}, \citenamefont {Sangouard}, \citenamefont
  {Stobi\ifmmode~\acute{n}\else \'{n}\fi{}ska}, \citenamefont {Bussi\`eres},
  \citenamefont {Afzelius},\ and\ \citenamefont {Gisin}}]{Sekatski}%
  \BibitemOpen
  \bibfield  {author} {\bibinfo {author} {\bibfnamefont {P.}~\bibnamefont
  {Sekatski}}, \bibinfo {author} {\bibfnamefont {N.}~\bibnamefont {Sangouard}},
  \bibinfo {author} {\bibfnamefont {M.}~\bibnamefont
  {Stobi\ifmmode~\acute{n}\else \'{n}\fi{}ska}}, \bibinfo {author}
  {\bibfnamefont {F.}~\bibnamefont {Bussi\`eres}}, \bibinfo {author}
  {\bibfnamefont {M.}~\bibnamefont {Afzelius}}, \ and\ \bibinfo {author}
  {\bibfnamefont {N.}~\bibnamefont {Gisin}},\ }\href {\doibase
  10.1103/PhysRevA.86.060301} {\bibfield  {journal} {\bibinfo  {journal} {Phys.
  Rev. A}\ }\textbf {\bibinfo {volume} {86}},\ \bibinfo {pages} {060301(R)}
  (\bibinfo {year} {2012})}\BibitemShut {NoStop}%
\bibitem [{\citenamefont {De~Martini}\ \emph {et~al.}(2008)\citenamefont
  {De~Martini}, \citenamefont {Sciarrino},\ and\ \citenamefont
  {Vitelli}}]{demartini}%
  \BibitemOpen
  \bibfield  {author} {\bibinfo {author} {\bibfnamefont {F.}~\bibnamefont
  {De~Martini}}, \bibinfo {author} {\bibfnamefont {F.}~\bibnamefont
  {Sciarrino}}, \ and\ \bibinfo {author} {\bibfnamefont {C.}~\bibnamefont
  {Vitelli}},\ }\href {\doibase 10.1103/PhysRevLett.100.253601} {\bibfield
  {journal} {\bibinfo  {journal} {Phys. Rev. Lett.}\ }\textbf {\bibinfo
  {volume} {100}},\ \bibinfo {pages} {253601} (\bibinfo {year}
  {2008})}\BibitemShut {NoStop}%
\bibitem [{\citenamefont {Iskhakov}\ \emph {et~al.}(2012)\citenamefont
  {Iskhakov}, \citenamefont {Agafonov}, \citenamefont {Chekhova},\ and\
  \citenamefont {Leuchs}}]{Iskhakov}%
  \BibitemOpen
  \bibfield  {author} {\bibinfo {author} {\bibfnamefont {T.~S.}\ \bibnamefont
  {Iskhakov}}, \bibinfo {author} {\bibfnamefont {I.~N.}\ \bibnamefont
  {Agafonov}}, \bibinfo {author} {\bibfnamefont {M.~V.}\ \bibnamefont
  {Chekhova}}, \ and\ \bibinfo {author} {\bibfnamefont {G.}~\bibnamefont
  {Leuchs}},\ }\href {\doibase 10.1103/PhysRevLett.109.150502} {\bibfield
  {journal} {\bibinfo  {journal} {Phys. Rev. Lett.}\ }\textbf {\bibinfo
  {volume} {109}},\ \bibinfo {pages} {150502} (\bibinfo {year}
  {2012})}\BibitemShut {NoStop}%
\bibitem [{\citenamefont {Stobi\ifmmode~\acute{n}\else \'{n}\fi{}ska}\ \emph
  {et~al.}(2012{\natexlab{a}})\citenamefont {Stobi\ifmmode~\acute{n}\else
  \'{n}\fi{}ska}, \citenamefont {T\"oppel}, \citenamefont {Sekatski},\ and\
  \citenamefont {Chekhova}}]{BSV}%
  \BibitemOpen
  \bibfield  {author} {\bibinfo {author} {\bibfnamefont {M.}~\bibnamefont
  {Stobi\ifmmode~\acute{n}\else \'{n}\fi{}ska}}, \bibinfo {author}
  {\bibfnamefont {F.}~\bibnamefont {T\"oppel}}, \bibinfo {author}
  {\bibfnamefont {P.}~\bibnamefont {Sekatski}}, \ and\ \bibinfo {author}
  {\bibfnamefont {M.~V.}\ \bibnamefont {Chekhova}},\ }\href {\doibase
  10.1103/PhysRevA.86.022323} {\bibfield  {journal} {\bibinfo  {journal} {Phys.
  Rev. A}\ }\textbf {\bibinfo {volume} {86}},\ \bibinfo {pages} {022323}
  (\bibinfo {year} {2012}{\natexlab{a}})}\BibitemShut {NoStop}%
\bibitem [{\citenamefont {Stobi\ifmmode~\acute{n}\else \'{n}\fi{}ska}\ \emph
  {et~al.}(2012{\natexlab{b}})\citenamefont {Stobi\ifmmode~\acute{n}\else
  \'{n}\fi{}ska}, \citenamefont {T\"oppel}, \citenamefont {Sekatski},
  \citenamefont {Buraczewski}, \citenamefont {\ifmmode~\dot{Z}\else
  \.{Z}\fi{}ukowski}, \citenamefont {Chekhova}, \citenamefont {Leuchs},\ and\
  \citenamefont {Gisin}}]{MDF}%
  \BibitemOpen
  \bibfield  {author} {\bibinfo {author} {\bibfnamefont {M.}~\bibnamefont
  {Stobi\ifmmode~\acute{n}\else \'{n}\fi{}ska}}, \bibinfo {author}
  {\bibfnamefont {F.}~\bibnamefont {T\"oppel}}, \bibinfo {author}
  {\bibfnamefont {P.}~\bibnamefont {Sekatski}}, \bibinfo {author}
  {\bibfnamefont {A.}~\bibnamefont {Buraczewski}}, \bibinfo {author}
  {\bibfnamefont {M.}~\bibnamefont {\ifmmode~\dot{Z}\else \.{Z}\fi{}ukowski}},
  \bibinfo {author} {\bibfnamefont {M.~V.}\ \bibnamefont {Chekhova}}, \bibinfo
  {author} {\bibfnamefont {G.}~\bibnamefont {Leuchs}}, \ and\ \bibinfo {author}
  {\bibfnamefont {N.}~\bibnamefont {Gisin}},\ }\href {\doibase
  10.1103/PhysRevA.86.063823} {\bibfield  {journal} {\bibinfo  {journal} {Phys.
  Rev. A}\ }\textbf {\bibinfo {volume} {86}},\ \bibinfo {pages} {063823}
  (\bibinfo {year} {2012}{\natexlab{b}})}\BibitemShut {NoStop}%
\bibitem [{\citenamefont {Buraczewski}\ and\ \citenamefont
  {Stobi\'nska}(2012)}]{CPC}%
  \BibitemOpen
  \bibfield  {author} {\bibinfo {author} {\bibfnamefont {A.}~\bibnamefont
  {Buraczewski}}\ and\ \bibinfo {author} {\bibfnamefont {M.}~\bibnamefont
  {Stobi\'nska}},\ }\href {\doibase 10.1016/j.cpc.2012.04.027} {\bibfield
  {journal} {\bibinfo  {journal} {Comp. Phys. Commun.}\ }\textbf {\bibinfo
  {volume} {183}},\ \bibinfo {pages} {2245} (\bibinfo {year}
  {2012})}\BibitemShut {NoStop}%
\bibitem [{\citenamefont {Stobińska}(2015)}]{Feasibility}%
  \BibitemOpen
  \bibfield  {author} {\bibinfo {author} {\bibfnamefont {M.}~\bibnamefont
  {Stobińska}},\ }\href {\doibase
  http://dx.doi.org/10.1016/j.optcom.2014.07.030} {\bibfield  {journal}
  {\bibinfo  {journal} {Optics Communications}\ }\textbf {\bibinfo {volume}
  {337}},\ \bibinfo {pages} {83 } (\bibinfo {year} {2015})},\ \bibinfo {note}
  {macroscopic quantumness: theory and applications in optical
  sciences}\BibitemShut {NoStop}%
\bibitem [{\citenamefont {Vitell}\ \emph {et~al.}(2009)\citenamefont {Vitell},
  \citenamefont {Spagnolo}, \citenamefont {Sciarrino},\ and\ \citenamefont
  {De~Martini}}]{production}%
  \BibitemOpen
  \bibfield  {author} {\bibinfo {author} {\bibfnamefont {C.}~\bibnamefont
  {Vitell}}, \bibinfo {author} {\bibfnamefont {N.}~\bibnamefont {Spagnolo}},
  \bibinfo {author} {\bibfnamefont {F.}~\bibnamefont {Sciarrino}}, \ and\
  \bibinfo {author} {\bibfnamefont {F.}~\bibnamefont {De~Martini}},\ }\href
  {\doibase 10.1364/JOSAB.26.000892} {\bibfield  {journal} {\bibinfo  {journal}
  {J. Opt. Soc. Am. B}\ }\textbf {\bibinfo {volume} {26}},\ \bibinfo {pages}
  {892} (\bibinfo {year} {2009})}\BibitemShut {NoStop}%
\bibitem [{\citenamefont {Spagnolo}\ \emph {et~al.}(2009)\citenamefont
  {Spagnolo}, \citenamefont {Vitelli}, \citenamefont {De~Angelis},
  \citenamefont {Sciarrino},\ and\ \citenamefont {De~Martini}}]{spagnolo}%
  \BibitemOpen
  \bibfield  {author} {\bibinfo {author} {\bibfnamefont {N.}~\bibnamefont
  {Spagnolo}}, \bibinfo {author} {\bibfnamefont {C.}~\bibnamefont {Vitelli}},
  \bibinfo {author} {\bibfnamefont {T.}~\bibnamefont {De~Angelis}}, \bibinfo
  {author} {\bibfnamefont {F.}~\bibnamefont {Sciarrino}}, \ and\ \bibinfo
  {author} {\bibfnamefont {F.}~\bibnamefont {De~Martini}},\ }\href {\doibase
  10.1103/PhysRevA.80.032318} {\bibfield  {journal} {\bibinfo  {journal} {Phys.
  Rev. A}\ }\textbf {\bibinfo {volume} {80}},\ \bibinfo {pages} {032318}
  (\bibinfo {year} {2009})}\BibitemShut {NoStop}%
\bibitem [{\citenamefont {Kr\'al}(1990)}]{Kral}%
  \BibitemOpen
  \bibfield  {author} {\bibinfo {author} {\bibfnamefont {P.}~\bibnamefont
  {Kr\'al}},\ }\href {\doibase 10.1080/09500349014550941} {\bibfield  {journal}
  {\bibinfo  {journal} {J. Mod. Optic}\ }\textbf {\bibinfo {volume} {37}},\
  \bibinfo {pages} {889} (\bibinfo {year} {1990})}\BibitemShut {NoStop}%
\bibitem [{\citenamefont {Clauser}\ \emph {et~al.}(1969)\citenamefont
  {Clauser}, \citenamefont {Horne}, \citenamefont {Shimony},\ and\
  \citenamefont {Holt}}]{CHSH}%
  \BibitemOpen
  \bibfield  {author} {\bibinfo {author} {\bibfnamefont {J.~F.}\ \bibnamefont
  {Clauser}}, \bibinfo {author} {\bibfnamefont {M.~A.}\ \bibnamefont {Horne}},
  \bibinfo {author} {\bibfnamefont {A.}~\bibnamefont {Shimony}}, \ and\
  \bibinfo {author} {\bibfnamefont {R.~A.}\ \bibnamefont {Holt}},\ }\href
  {\doibase 10.1103/PhysRevLett.23.880} {\bibfield  {journal} {\bibinfo
  {journal} {Phys. Rev. Lett.}\ }\textbf {\bibinfo {volume} {23}},\ \bibinfo
  {pages} {880} (\bibinfo {year} {1969})}\BibitemShut {NoStop}%
\bibitem [{\citenamefont {Nielsen}\ and\ \citenamefont
  {Chuang}(2000)}]{NielsenChuangBook}%
  \BibitemOpen
  \bibfield  {author} {\bibinfo {author} {\bibfnamefont {M.~A.}\ \bibnamefont
  {Nielsen}}\ and\ \bibinfo {author} {\bibfnamefont {I.~L.}\ \bibnamefont
  {Chuang}},\ }\href@noop {} {\emph {\bibinfo {title} {Quantum Computation and
  Quantum Information}}},\ \bibinfo {edition} {1st}\ ed.\ (\bibinfo
  {publisher} {Cambridge University Press},\ \bibinfo {year}
  {2000})\BibitemShut {NoStop}%
\bibitem [{\citenamefont {Yurke}\ and\ \citenamefont
  {Stoler}(1987)}]{PhysRevA.36.1955}%
  \BibitemOpen
  \bibfield  {author} {\bibinfo {author} {\bibfnamefont {B.}~\bibnamefont
  {Yurke}}\ and\ \bibinfo {author} {\bibfnamefont {D.}~\bibnamefont {Stoler}},\
  }\href {\doibase 10.1103/PhysRevA.36.1955} {\bibfield  {journal} {\bibinfo
  {journal} {Phys. Rev. A}\ }\textbf {\bibinfo {volume} {36}},\ \bibinfo
  {pages} {1955} (\bibinfo {year} {1987})}\BibitemShut {NoStop}%
\bibitem [{\citenamefont {Gardiner}\ and\ \citenamefont
  {Zoller}(2004)}]{GardinerZollerBook}%
  \BibitemOpen
  \bibfield  {author} {\bibinfo {author} {\bibfnamefont {C.~W.}\ \bibnamefont
  {Gardiner}}\ and\ \bibinfo {author} {\bibfnamefont {P.}~\bibnamefont
  {Zoller}},\ }\href@noop {} {\emph {\bibinfo {title} {Quantum Noise}}},\
  \bibinfo {edition} {3rd}\ ed.\ (\bibinfo  {publisher} {Springer},\ \bibinfo
  {year} {2004})\BibitemShut {NoStop}%
\bibitem [{Note1()}]{Note1}%
  \BibitemOpen
  \bibinfo {note} {\protect \leavevmode {\protect \color {black}If the
  two-photon N00N state is not amplified at the source but instead transmitted
  to Alice and Bob before amplification, we find $t^\protect \mathrm
  {min}=79.1\%$.}}\BibitemShut {Stop}%
\bibitem [{\citenamefont {Zavatta}\ \emph {et~al.}(2004)\citenamefont
  {Zavatta}, \citenamefont {Viciani},\ and\ \citenamefont {Bellini}}]{Zavatta}%
  \BibitemOpen
  \bibfield  {author} {\bibinfo {author} {\bibfnamefont {A.}~\bibnamefont
  {Zavatta}}, \bibinfo {author} {\bibfnamefont {S.}~\bibnamefont {Viciani}}, \
  and\ \bibinfo {author} {\bibfnamefont {M.}~\bibnamefont {Bellini}},\ }\href
  {\doibase 10.1103/PhysRevA.70.053821} {\bibfield  {journal} {\bibinfo
  {journal} {Phys. Rev. A}\ }\textbf {\bibinfo {volume} {70}},\ \bibinfo
  {pages} {053821} (\bibinfo {year} {2004})}\BibitemShut {NoStop}%
\end{thebibliography}%

\end{document}